# DNA looping and physical constraints on transcription regulation

José M. G. Vilar[*] and Stanislas Leibler

The Rockefeller University, 1230 York Avenue, Box 34, New York, NY 10021

[*] e-mail: `vilarj@rockefeller.edu`

## Abstract

DNA looping participates in transcriptional regulation, for instance, by allowing distal binding sites to act synergistically. Here we study this process and compare different regulatory mechanisms based on repression with and without looping. Within a simple mathematical model for the *lac* operon, we show that regulation based on DNA looping, in addition to increasing the repression level, can reduce the fluctuations of transcription and, at the same time, decrease the sensitivity to changes in the number of regulatory proteins. Looping is thus able to circumvent some of the constraints inherent to mechanisms based solely on binding to a single operator site and provides a mechanism to regulate not only the average properties of transcription but also its fluctuations.

# Introduction

Cells use a wide variety of mechanisms to regulate and perform their functions. Some of these mechanisms are fairly simple. But, more often that not, there seems to be an unnecessary complexity. Consider for instance the *lac* operon, the system where, together with $\lambda$-phage, gene regulation was discovered.[1,2,3] It consists of a regulatory domain and three genes required for the uptake and catabolism of lactose (see Figure 1). A regulatory protein, the *lac* repressor, can bind to the main operator $O_1$ and prevent the RNA polymerase from transcribing the genes. If it is not bound, transcription proceeds at a given rate. This simple idea emerged as one of the milestones of gene regulation: there are DNA binding proteins that can hinder or stimulate some of the steps leading to transcription. Regulation of transcription, however, is actually not so simple. In the case of the *lac* operon, besides $O_1$ there are two sites outside the control region, the so-called auxiliary operators $O_2$ and $O_3$, which closely resemble $O_1$ and where the repressor can also bind. At first, these two sites were considered to be just remnants of evolution without any specific function.[2] The reasons were diverse. They are far away from the promoter, so that the repressor's binding to them cannot affect the RNA polymerase directly. They are much weaker than $O_1$ — $O_2$ is as much as 10 times and $O_3$ is over 300 times. Moreover, elimination of either one of them leaves the repression level practically unchanged.

The role of $O_2$ and $O_3$, however, proved to be far from minor: simultaneous elimination of both of these operators reduced the repression level about 100 times.

Such a drastic effect turned out to be mediated by the DNA loops that the *lac* repressor can induce by binding to two sites simultaneously.[4] Through looping, the auxiliary operators indirectly increase the probability for the repressor to be bound to the main operator.

It is remarkable that, despite its apparent complexity, DNA looping is widely used in gene regulation. It was first discovered in the *ara* operon[5] and subsequently, in other prokaryotic systems like *lac*, *deo*, *gal* and *gln*.[6] It is a key element in the regulation of the $\lambda$–phage[7] and it is also at play in eukaryotic enhancers, allowing multiple proteins from adjacent and also distal sites to affect the RNA polymerase.[6,8]

Here we analyze how the dynamics of looping affects gene expression and compare it to different alternative regulatory mechanisms. The results of our model are in close agreement with the available experimental data on the *lac* operon, which spans over three orders of magnitude in the repression level. In addition, the model shows that DNA looping can be used to circumvent several of the shortcomings that are inherent to simpler mechanisms.

## Regulation with and without looping

In Figure 2 we illustrate the main differences in the mechanisms of regulation with and without looping. The system with a single binding site can be characterized by two states (Figure 2a). In the state (*i*) the operator $O_m$ is not occupied and in the state (*ii*) one of the $N$ repressors of the cell is bound to $O_m$. In principle, one might think of a more detailed description of the system, e.g. including states for the repressor bound non-specifically to DNA or freely diffusing in the cell. Such a detailed description would result at the end in an effective two-state description. Here, states are chosen to keep just the essential elements.

The DNA looping case is more complex and interesting (Figure 2b). The major contribution of looping to gene regulation comes from the synergistic effects of two operators. Thus, we consider the two-operator case, for which there exists the most detailed experimental data.[14] Now, there are five relevant states: (*i*) none of the operators is occupied, (*ii*) a repressor is bound to just the main operator $O_m$, (*iii*) to just the auxiliary operator $O_a$, (*iv*) to both of the operators by looping DNA, or (*v*) one repressor is bound to $O_m$ and other to $O_a$ at the same time.

**Repression level**

The description based on states is suitable to tackle, both qualitatively and quantitatively, the effects of looping in gene regulation. Intuitively, looping increases repression because the system is dynamically trapped in the looped state (*iv*). The system can only leave this state to either the state (*ii*) or (*iii*). In any of these two states, the repressor still remains nearby the free operator. Therefore, the most likely event is that repressor is recaptured by the free operator to form the loop again. Thus, with high probability, the system comes back to the state (*iv*).

This idea of the repressor being dynamically trapped is also a key element in a recently proposed mechanism for protein localization.[9] Proteins with two binding domains for each of the elements of an array will have a high probability of being attached to the array by one or both of its domains at any instant of time because, if the neighboring array elements are close enough, it is likely that when one domain unbinds it will reattach to the array before the other domain unbinds.

It is important to note the differences of DNA looping with what is known in inorganic chemistry as the chelate effect.[10] The chelate effect refers to the fact that the binding of a dimer to a molecule may be far greater than expected from the binding of the constituent monomers separately. It happens because, in the binding, the dimer loses only the translational and rotational entropies of a single molecule in contrast to the entropies of two molecules that the pair of monomers would lose. In our case, in addition to the lost of translational and rotational entropies, one also has to take into account the energetic and entropic contribution of the formation of the DNA loop.

To proceed with the quantitative details, we consider first the single operator case. We will follow the standard statistical thermodynamics approach.[11] The main idea of this approach is that the probability for the system to be in a given state is a function of the free energy of such a state. This function is essentially proportional to the number of ways in which the state can be realized times the exponential of minus the free energy of the state.[11,12,13] From these probabilities, one can obtain all the equilibrium properties of the system. In our case, the quantity of interest is the repression level $R_{O_m}$, which is defined as the ratio of the maximum transcription rate ($t_{max}$) to the actual rate ($t_{act}$). If transcription takes place when the repressor is not bound to the main operator, as in our case, the actual transcription rate is the maximum rate times the probability for the main operator to be free. The main operator is free when the system is in the state ($i$). Therefore, $R_{O_m} \equiv t_{max}/t_{act} = t_{max}/(t_{max} P_i)$ leads to

$$R_{O_m} = \frac{1}{P_i} = 1 + \frac{P_{ii}}{P_i} = 1 + N e^{-\Delta G_{O_m}}, \qquad (1)$$

where $P$ is the probability for the system to be in the state denoted by its subscript, $\Delta G_{O_m}$ is the change in free energy due to the binding of the repressor to $O_m$ (in units of $k_B T$, with $k_B$ the Boltzmann constant and $T$ the absolute temperature) and $N$ is the number of repressors. The factor $N$ appears in Equation (1) because one out of $N$ repressors can bind to the site — therefore, there are $N$ possible ways in which the state can be realized. Notice also that we have used the probability normalization condition ($P_i + P_{ii} = 1$). An interesting aspect of this mathematical analysis is that it relates the physical (free energy) to the genetic (repression level) properties of the system.

There is a subtle but conceptually important difference with the most common biochemical approach: we use the number of repressors per cell instead of concentrations. Thinking in terms of concentrations is useful for analyzing *in vitro* experiments, but concentrations might not be well defined in a heterogeneous, non-uniform and crowded environment like the interior of the cell. Once the repression level and the number of repressors are known from experiments,[14] the only unknown in Equation (1) is the free energy. By solving the resulting equation for different sets of parameters, we obtained the *in vivo* free energies for the binding of the repressor to $O_1$, $O_2$, and $O_3$ (see Table 1). The free energies obtained in this way are not just what is usually measured *in vitro*; they also take into account the nonspecific binding of the repressor and the looping between the main operator and nonspecific DNA (see Appendix B); i.e., they also take into account the context of the cell.

In a similar way, one can compute the repression level when DNA looping is involved. As before, the repression level is the inverse of the probability for the main operator to be free, which takes place when the system is in the states (*i*) and (*iii*). Therefore,

$$R_{O_m} = \frac{1}{P_i + P_{iii}} = 1 + \frac{P_{ii} + P_{iv} + P_v}{P_i + P_{iii}}. \tag{2}$$

In terms of free energies, we obtain

$$R_{O_m} = 1 + \frac{Ne^{-\Delta G_{O_m}} + Ne^{-(\Delta G_{O_m} + \Delta G_{O_a} + \Delta G_l)} + N(N-1)e^{-(\Delta G_{O_m} + \Delta G_{O_a})}}{1 + Ne^{-\Delta G_{O_a}}}, \quad (3)$$

where $\Delta G_{O_a}$ is the free energy of the repressor bound to the auxiliary operator $O_a$ and $\Delta G_l$ is the contribution of looping to the free energy (the free energy of the state (*iv*) is given by $\Delta G_{iv} \equiv \Delta G_{O_m} + \Delta G_{O_a} + \Delta G_l$).

Once the number of repressors per cell and the binding free energies for each site are known, $\Delta G_l$ can be obtained directly from Equation (3). This quantity depends on the particular experimental situation, mostly, on the distance between operators and on the elastic properties of DNA. It is expected not to depend on the strength of the different operators. The experiments of reference 14 measured the repression levels for the case in which the auxiliary operator located 92 bp upstream was deleted; i.e., only the main operator and the auxiliary operator located 401 bp downstream were present. The experiments were performed for two different numbers of repressors per cell and, in both cases, for three different sequences of the main operator (namely, for the sequences of $O_1$, $O_2$, and $O_3$). The sequence of the auxiliary operator was always the sequence of $O_2$. (For experimental details of reference 14 see the caption of Table 1.)

Using these experimental data and Equation (3), we calculated $\Delta G_l$ for each case (see Table 2). As free energies for the binding of the repressor to $O_1$, $O_2$, and $O_3$ we have used the average values of Table 1, which were obtained from the experimental data of reference 14. To check the consistency of the results, we have evaluated $R_{O_m}$ for a single value of $\Delta G_l$, taken to be the average $\langle \Delta G_l \rangle$ of the six values displayed in Table 2. The results obtained in this way are in close agreement with the experimental ones, as shown in Figure 3. This figure shows that, indeed,

Equation (3) properly accounts for the repression level over the three orders of magnitude of the experimental data. In general, $\Delta G_l$ will depend on the distance and on the sequence between operators.[16,12] All the experimental data we have used[14] kept these parameters constant, which is revealed in our approach by the relatively constant values of $\Delta G_l$ that were obtained.

## Increase in the local concentration

It is interesting to analyze how previous explanations of the effect of the auxiliary operators relate to Equation (3). To this end, we need to consider the case in which the auxiliary operator is sufficiently strong. In mathematical terms, this can be expressed as $Ne^{-\Delta G_{O_a}} \gg 1$. Under such conditions, Equation (3) simplifies to

$$R_{O_m} = 1 + \left[ e^{-\Delta G_l} + (N-1) \right] e^{-\Delta G_{O_m}} , \qquad (4)$$

where the repression level is as for the single operator case [Equation (1)] but now with an effective number of repressors per cell given by $N_{eff} = e^{-\Delta G_l} + (N-1)$. The effect of the auxiliary operator is thus analogous to increasing the number of repressors per cell.

The connection with the increase in the local concentration[15] explanation comes from assuming that when the repressor is bound to the auxiliary operator the sole effect of looping is to reduce the volume in which the repressor can move around the main operator. This is, of course, a very crude approximation, which nevertheless is able to provide some insights. Under this assumption, the decrease in free energy because of looping is given by $\Delta G_l = -\ln \frac{V_{cell}}{V_{loop}}$, where $V_{cell}$ is the volume of the cell and $V_{loop}$ is the volume in which the repressor is allowed to move once it is bound to the auxiliary site.[6,11] The repression level follows from Equation (1) with $N_{eff} = \frac{V_{cell}}{V_{loop}} + (N-1)$, which coincides with the result obtained from the increase in the local concentration.[15] Notice, however, that in general, repression through looping extends beyond the concept of local concentration, as shown by Equation (3).

## Dynamics and fluctuations

To study the fluctuations in the numbers of protein and mRNA molecules, the dynamics has to be considered explicitly. The new quantities of interest are the transition probability rates between different states. Unfortunately, so far, there is no *in vivo* measurement of those rates and the *in vitro* data strongly differs from experiment to experiment.[17] We estimated the rates as explained in the Appendix A, taking into account as much *in vivo* information as possible.

In Figure 4 we show the typical time courses and the histograms of the number of molecules produced from operons regulated with and without looping. These graphs were obtained from computer simulations using the standard Gillespie algorithm[18] (known as the BKL algorithm in the physical literature[19]). The basic idea of the algorithm is to choose randomly (with probabilities inferred from the rates[20]) a transition and the time at which it happens. The state of the system is updated accordingly and the procedure is repeated until some final state, or time limit, is reached. (For more details about the algorithm see references 18 and 19.)

Figure 4a corresponds to regulation with looping. In principle, there exist simple mechanisms involving a single operator that would result in the same repression level. For instance, one could use a stronger operator with a lower dissociation rate (by decreasing the off-rate of the repressor), as shown in Figure 4b, or, alternatively, a repressor with a higher association rate (by increasing the on-rate), as shown in Figure 4c. In both cases, the parameters of the system can be chosen so that the number of molecules fluctuates around the same average. Note, however, that the fluctuations around the average value are very different.

Figure 4 clearly illustrates that mechanisms that give the same repression level are not necessarily equivalent. For instance, eliminating the auxiliary operator and increasing the strength of the main operator (while keeping the same repression level) would result in higher fluctuations in mRNA and protein production. In contrast, increasing repression by increasing the association rate has basically the same fluctuations as the original system with DNA looping.

The reason for these differences is a matter of time scales. The rule of thumb is that the faster the fluctuations (i.e., the shorter their correlation time), the smaller the effect of the fluctuations is. The underlying idea is that a higher number of uncorrelated events per unit of time will result in a better average at longer time scales. This can be seen explicitly, for instance, in linear systems with correlated Gaussian noise[20] and in certain classes of non-linear systems.[21] Exceptions to this rule can appear when changes in the correlation time induce also other effects in the dynamics of the system. In our case, looping introduces a fast time scale: the time for the repressor to be recaptured by the main operator before unbinding the auxiliary operator. If transcription switches slowly between active and inactive, there are long periods of time in which proteins are produced constantly and long periods without any production. Therefore, the number of molecules fluctuates strongly between high and low values. In contrast, if the switching is very fast, the production is in the form of short and frequent bursts. This lack of long periods of time with either full or null production gives a narrower distribution of the number of molecules.

At a glance, looping and a higher association rate seem thus to provide equivalent mechanisms regarding the repression level and the fluctuations. There are, however, certain limits for the values that the rate constants can achieve. The theoretical limit for diffusion limited reactions[22] is about $k_a \simeq 10^9 M^{-1} s^{-1}$. The values inferred from our analysis and the experimental data on the *lac* operon (see Appendix A and Figure 4) are consistent with this upper limit. To reduce the fluctuations by increasing the association rate constant, the diffusion limit would have to be surpassed —the values used in Figure 4c are larger than this limit. Looping provides the cell with a mechanism to circumvent the physical constraints of diffusion limited reactions. Note that other mechanisms have been hypothesized for regulatory proteins to locate their targets at higher rates than those allowed by the diffusion limit, such as sliding on the DNA strand.[23]

## Cell-to-cell variability

The number of repressors is expected to differ from cell to cell. An important property is therefore the dependence of the repression level [Equation (3)] on the number of repressors. In Figure 5a we illustrate this dependence for both the looping and the single operator cases.

For a single operator, the repression level is a linear function of the number of repressors, i.e. a constant term plus a term proportional to the number of repressors. The proportionality factor is the repressor-operator equilibrium binding constant. In the looping case, the repression level is no longer a linear function of the number of repressors. Therefore, it is not possible to understand the looping case in terms of a single site with an increased effective binding constant. Interestingly, this nonlinear dependence of the repression level decreases the sensitivity of the repression level to variations in the number of repressors. It is then possible to attain fairly constant repression levels even in the presence of marked variable numbers of regulatory proteins, as Figure 5a clearly illustrates.

Even more interesting is the possibility to control not only the repression level but also its variability from cell to cell. Looping could in principle be used to adjust (over evolutionary time scales) the levels of phenotypic variability: the strength of the main and auxiliary operators as well as the distance between them could be chosen so that an optimal cell-to-cell variability of the repression level is obtained. In Figure 5b we show an instance of how cell-to-cell variability could be controlled by looping.

## Conclusions

The complexity of the cell contrasts with the simplicity of the idealized models aimed at its understanding. In the cell, the numbers of each molecular component are limited and often fluctuate strongly, not only in time but also from cell to cell. In addition, the reactions between components cannot happen at arbitrarily high speeds. The ability to cope with, integrate, and use these constraints is crucial for the functioning of the cell. Here we have studied how these constraints affect gene regulation. In particular, we have focused on the role of DNA looping, which seems to exhibit an unnecessary complexity when compared to alternative, apparently simpler mechanisms. A well established role of DNA looping is to increase the repression level. In principle, it would be also possible to increase the repression level by just increasing the strength of the operator or the affinity of repressor for the operator.

The results of our analysis suggest that DNA looping, in addition to increasing the repression level can confer other relevant properties to gene regulation systems:

– Compared to simpler alternative regulatory mechanisms, DNA looping is able to reduce the fluctuations in transcription.

– The experimental data seems to indicate that the search of repressor for its target has reached the limits that diffusion imposes to reaction rates; in order for the repressor to find its target faster, this limit would have to be surpassed. DNA looping can circumvent this constraint.

– DNA looping also makes the repression level remain fairly constant with respect to changes in the number of repressors.

It is important to realize that noise and fluctuations are ubiquitous at the molecular level. The cellular function has to be carried out under such conditions. Regulation systems have evolved to cope with all the constraints that the intrinsic molecular nature of the cell imposes. Uncovering the way in which it is achieved is of fundamental importance for understanding both naturally occurring and artificially designed[24,25] cellular systems.

## Appendix A: Transition rates

The transition rates between the different states are basically the association rate constant, $k_a$, which gives the rate for the repressor to find an operator, and the repressor-operator dissociation rate constants, $k_{O_m}$ and $k_{O_a}$. If the repressor is already bound to one operator the association rate constant for the other operator changes by a factor $a$ to $a \times k_a$. Similarly, the dissociation rate constants change to $b \times k_{O_m}$ and $b \times k_{O_a}$ when the repressor is bound to both operators simultaneously. The factor $b$ takes into account the energetic contribution of looping and the factor $a$, the entropic effects.

The binding of the repressor takes place under equilibrium conditions, which imposes additional constraints on the values the biochemical parameters can take. At equilibrium, the probability per unit of time of going from a given state $X$ to another state $Y$ is the same as that of going from $Y$ to $X$. This fact is known in statistical mechanics as the principle of detailed balance. In our case, it implies that

$$ak_a P_{ii} = bk_{O_a} P_{iv}, \quad (5)$$

$$ak_a P_{iii} = bk_{O_m} P_{iv}, \quad (6)$$

$$k_a N P_i = k_{O_m} P_{ii}, \quad (7)$$

$$k_a N P_i = k_{O_a} P_{iii}, \quad (8)$$

which together with the equilibrium probabilities $\frac{P_{ii}}{P_i} = Ne^{-\Delta G_{O_m}}$, $\frac{P_{iii}}{P_i} = Ne^{-\Delta G_{O_a}}$, $\frac{P_{iv}}{P_i} = Ne^{-(\Delta G_{O_m}+\Delta G_{O_a}+\Delta G_l)}$, and $\frac{P_v}{P_i} = N(N-1)e^{-(\Delta G_{O_m}+\Delta G_{O_a})}$ leads to $a = be^{-\Delta G_l}$, $k_a = k_{O_m} e^{-\Delta G_{O_m}}$, and $k_a = k_{O_a} e^{-\Delta G_{O_a}}$.

Therefore, if $b$ (or alternatively $a$) and one rate constant are known, all the others follow from the free energies for the different states. So far, there is no direct measurement of the *in vivo* values of those constants. The *in vitro* data shows a high variability —as much as 100-fold differences— from experiment to experiment.[17] As a value of the dissociation rate constant for $O_1$ we have chosen $k_{O_1} = 0.016\, s^{-1}$, according to experiments that used a short piece of DNA with just the operator.[26] (Other experiments used long pieces of DNA, which can induce looping and actually not provide the dissociation rate constant from a single operator.[17])

As a value of $b$, we have assumed $b = 1$, which means that the dissociation rate of the repressor from one operator does not depend on whether it is also bound to the other operator. This is reasonable on the grounds that the two operators are far apart, beyond the DNA persistence length. If DNA elastic, torsional, and repulsive effects are relevant and dominate over the attractive ones, one would have $b>1$. On the other hand, if attractive effects dominate, one would have $b<1$. Irrespective of the particular value of $b$, looping would reduce the transcriptional noise provided that $a>1$. In general, a bigger $b$ implies a bigger $a$ and, therefore, a more pronounced reduction of the transcriptional noise.

With the values of $b$ and $k_{O_1}$, the principle of detailed balance, and the data from Tables 1 and 2, we can infer the values of all the other constants.

## Appendix B: Nonspecific binding and DNA looping

One issue that arises when choosing the states of the system is the level of detail one should consider. For instance, in the binding of the repressor to a single operator we considered that there are only two states (see Figure 2a). The repressor is either bound [state (*ii*)] or not bound [state (*i*)] to the operator. In fact, because of nonspecific binding, the state (*ii*) can be considered as composed of several sub-states. In such a state, one binding domain of the repressor is always bound to the operator but the other domain can be either free or bound to nonspecific DNA forming a DNA loop.

We label the sub-state without nonspecific DNA looping by (*ii,0*), and the sub-states with it, by (*ii,k*). Here, *k* is an index ranging from 1 to *n*, with *n* being the number of possible non-specific DNA binding sites. Then, proceeding as for Equation (1), one obtains that the repression level is given by

$$R_{O_m} = 1 + \frac{P_{ii,0} + \sum_{k=1}^{k=n} P_{ii,k}}{P_i}, \tag{9}$$

which expressed in terms of free energies leads to

$$R_{O_m} = 1 + Ne^{-\Delta \tilde{G}_{O_m}} + N \sum_{k=1}^{k=n} e^{-(\Delta \tilde{G}_{O_m} + \Delta \tilde{G}_{O_k} + \Delta \tilde{G}_{l_k})}, \tag{10}$$

where, $\Delta \tilde{G}_{O_m}$ is the change in free energy when one domain of the repressor binds to the operator while the other domain is free; $\Delta \tilde{G}_{O_k}$ is the same as $\Delta \tilde{G}_{O_m}$ but refereed to the binding to nonspecific DNA at the site labeled by $k$; and $\Delta \tilde{G}_{l_k}$ is the corresponding looping contribution.

The key idea in the simplification process is noticing that the repression level can be rewritten as

$$R_{O_m} = 1 + Ne^{-\Delta G_{O_m}}, \tag{11}$$

with

$$\Delta G_{O_m} = \Delta \tilde{G}_{O_m} + \ln\left(1 + \sum_{k=1}^{k=n} e^{-(\Delta \tilde{G}_{O_k} + \Delta \tilde{G}_{l_k})}\right). \tag{12}$$

This result coincides with that of Equation (1). Therefore, the two-state description of Figure 2a is equivalent to the more involved description that considers non-specific DNA looping. The advantage of using the simple over the complex description is that, in our case, the relevant parameters are not $\Delta \tilde{G}_{O_m}$, $\Delta \tilde{G}_{O_k}$, and $\Delta \tilde{G}_{l_k}$ ($0 < k \leq n$), but their combination into a single effective free energy $\Delta G_{O_m}$, which can be inferred from physiological experiments that measure the repression level.

## Acknowledgements

We thank C. Guet, B. Müller-Hill, M. Ptashne, and W. Shou for discussions and bibliographycomments on the manuscript. This work was supported in part by the U.S. National Institutes of Health.

# References

1. Jacob, F. & Monod, J. (1961). Genetic regulatory mechanisms in the synthesis of proteins. *J. Mol. Biol.* **3**, 318–356.

2. Müller-Hill, B. (1996). The lac Operon: A Short History of a Genetic Paradigm (Walter De Gruyter, Berlin).

3. Ptashne, M. (1992). A Genetic Switch: Phage λ and higher organisms (Blackwell Science, Cambridge).

4. Oehler, S., Eismann, E.R., Kramer, H. & Müller-Hill., B. (1990). The three operators of the lac operon cooperate in repression. *EMBO J.* **9**, 973–979.

5. Dunn, T. M., Hahn, S., Ogden, S. & Schleif, R.F. (1984). An operator at -280 base pairs that is required for repression of araBAD operon promoter: addition of DNA helical turns between the operator and promoter cyclically hinders repression. *Proc. Natl. Acad. Sci. U.S.A.* **81**, 5017–5020.

6. Schleif, R. (1992). DNA Looping. *Annu. Rev. Biochem.* **61**, 199–223.

7. Hochschild, A. (2002). The lambda switch: cl closes the gap in autoregulation. *Curr. Biol.* **12**, R87–R89.

8. Ptashne, M. & Gann, A. (2002) Genes & Signals (Cold Spring Harbor Laboratory Press, New York).

9. Levin, M.D., Shimizu, T.S. & Bray, D. (2002). Binding and diffusion of CheR molecules within a cluster of membrane receptors. *Biophys. J.* **82**, 1809–1817.

10. Page, M. I. & Jencks, W. P. (1971). Entropic Contributions to Rate Accelerations

in Enzymic and Intramolecular Reactions and the Chelate Effect. *Proc. Natl. Acad. Sci. U.S.A.* **68**, 1678–1683.

11. Hill, T. L. (1986). An Introduction to Statistical Thermodynamics (Dover, New York).

12. Law, S.M., Bellomy, G.R., Schlax, P.J. & Record, M.T. Jr. (1993). In Vivo Thermodynamic Analysis of Repression with and without Looping in lac Constructs. *J. Mol. Biol.* **230**, 161–173.

13. Ackers, G.K., Johnson, A.D. & Shea, M.A. (1982). Quantitative Model for Gene Regulation by λ Phage Repressor. *Proc. Natl. Acad. Sci. U.S.A.* **79**, 1129–1133.

14. Oehler, S., Amouyal, M., Kolkhof, P., von Wilcken-Bergmann, B. & Müller-Hill, B. (1994). Quality and position of the three lac operators of E. coli define efficiency of repression. *EMBO J.* **13**, 3348–3355.

15. Müller-Hill, B. (1998). The function of auxiliary operators. *Mol. Microbiol* **29**, 13–18.

16. Müller, J., Oehler, S. & Müller-Hill, B. (1996). Repression of lac promoter as a function of distance, phase and quality of an auxiliary lac operator. *J. Mol. Biol.* **257**, 21–29.

17. Barkley, M.D. & Bourgeois, S. (1980). Repressor recognition of operator and effectors. In Miller, J.H. and Reznikoff, W.S. (eds), The Operon (Cold Spring Harbor Laboratory Press, Cold Spring Harbor, NY) pp. 177–220.

18. Gillespie, D.T. (1977). Exact stochastic simulation of coupled chemical reactions. *J. Phys. Chem.* **81**, 2340–2361.

19. Bortz, A.B., Kalos, M.H. & Lebowitz, J.L. (1975). A New Algorithm for Monte Carlo Simulation of Ising Spin Systems. *J. Comp. Phys.* **17**:10–18.

20. van Kampen, N.G. (1981) Stochastic processes in physics and chemistry (North-

Holland, New York).

21. Vilar, J.M.G. & Rubí, J.M. (2001). Noise suppression by noise. *Phys. Rev. Lett.* **86**, 950–953.

22. Berg, O.G. & von Hippel, P.H. (1985). Diffusion-Controlled Macromolecular Interactions. *Annu. Rev. Biophys. Biophys. Chem.* **14**, 131–60.

23. Berg, O.G., Winter, R.B. & von Hippel, P.H. (1982). How do genome-regulatory proteins locate their DNA target sites? *Trends Biochem. Sci.* **7**,52–55.

24. Guet, C.C., Elowitz, M.B., Hsing, W.H. & Leibler, S. (2002). Combinatorial synthesis of genetic networks. *Science* **296**, 1466–1470.

25. Elowitz, M.B. & Leibler, S. (2000). A synthetic oscillatory network of transcriptional regulators. *Nature* **403**, 335–338.

26. Goeddel, D.V., Yansura, D.G. & Caruthers, M.H. (1977). How lac Repressor Recognizes lac Operator. *Proc. Natl. Acad. Sci. U.S.A.* **74**, 3292–3296.

# Figure Captions

**FIGURE 1:** Schematic representation of the *lac* operon (not drawn to scale). The three genes *lacZ*, *lacY*, and *lacA* are cotranscribed as a polycistronic message from a single promoter. The gene *lacZ* encodes for the β-galactosidase; *lacY*, for the permease; and *lacA*, for the transacetylase. The *lac* repressor is encoded by *lacI*, which is immediately upstream the operon. Binding of the repressor to the main operator site $O_1$ prevents transcription. The repressor can also bind to the auxiliary operators, $O_2$ and $O_3$. There is also an activator site, A, where the CAP-cAMP complex must bind for significant transcription.

**FIGURE 2:** Representative states of the binding of the repressor to **(a)** one and **(b)** two operators. Transcription takes place only in the states (*i*) and (*iii*), when $O_m$ is not occupied. The arrows indicate the possible transitions between states. Note that in (a) a single unbinding event is enough for the repressor to completely leave the neighborhood of the main operator. In (b) the repressor can escape from the neighborhood of the main operator only if it unbinds sequentially both operators. This sequence of events is highly unlikely for the typical values of the rate constants.

**FIGURE 3:** Computed repression (see Table 2) as a function of the observed repression.[14] The continuous line is the identity function.

**FIGURE 4:** Time series **(left)** and histograms **(right)** of the number of molecules

produced from operons regulated with and without looping. When the repressor is not bound to $O_m$, molecules are randomly produced at rate of 6.67 per second and degraded at a constant rate so that their characteristic life time is 30 min. **(a)** Regulation through looping with $k_{O_1} = 0.016\ s^{-1}$, $k_a = k_{O_1} e^{-\Delta G_{O_1}} = 0.073\ s^{-1}$, $k_{O_2} = 0.19\ s^{-1}$, $k_{O_3} = 7.33\ s^{-1}$, $b=1$, $a = b \times e^{-\Delta G_l} = 596$, and $N=10$. **(b)** Regulation with a single operator for the value of the repressor-operator dissociation rate constant, $k_{O_1} = 0.021 \times 0.016\ s^{-1}$, chosen so that the repression level is the same as in (a). The association rate constant remains unchanged. **(c)** Same situation as in (b) but now with $k_{O_1} = 0.016\ s^{-1}$ and $k_a = (1/0.021) \times 0.073\ s^{-1}$. Note that the association rate constants $k_a$ are per molecule. If we assume that the cellular volume is $2 \times 10^{-15}$ liters, we can express these constants in terms of concentrations: $k_a = 8.8 \times 10^7\ M^{-1} s^{-1}$ for (a) and (b) and $k_a = 4.1 \times 10^{10}\ M^{-1} s^{-1}$ for (c). This last value of $k_a$ is over the limit of diffusion-limited reactions (see main text).

**FIGURE 5:** **(a)** Repression level for the looping (continuous line) and single operator (dashed line) cases as a function of the number of repressors per cell. The continuous line was obtained from Equation (3) with $\Delta G_{O_m} = \langle \Delta G_{O_1} \rangle$, $\Delta G_{O_a} = \langle \Delta G_{O_2} \rangle$, and $\Delta G_l = \langle \Delta G_l \rangle$ from Tables 1 and 2. The dashed line was obtained from Equation (1) with $\Delta G_{O_m} = \langle \Delta G_{O_1} \rangle - \ln(0.021)$. The term $-\ln(0.021)$ was added to obtain the same repression level as in the looping case for $N \simeq 10$. This value of $N$ corresponds to the wild type average number of repressors per cell. The vertical bars indicate the size of the fluctuations in the repression level for the looping (black bar) and single operator (gray bar) cases that would arise as a result of fluctuations of the size of the horizontal bar in the number of repressors. **(b)** Cell-to-cell variability in the repression level as a result of variability in the number of repressors per cell. The color of each cell has been selected from Equation (4) for a random $N$ obtained from a Gaussian distribution with mean 10 and standard deviation 5. White, black, and gray correspond to high, low, and intermediate repression levels, respectively. **(left panel)** No looping: $\Delta G_{O_m} = -\ln(100)$ and $\Delta G_l = -0$. **(center panel)** Weak looping contribution: $\Delta G_{O_m} = -\ln(50)$ and $\Delta G_l = -\ln(11)$. **(right panel)** Strong looping contribution: $\Delta G_{O_m} = 0$ and $\Delta G_l = -\ln(991)$. The values of the free energies have been chosen so that the repression level is the same at $N=10$ for all three cases.

# Table Captions

**TABLE 1:** Free energies, $\Delta G_{O_m}$, of the binding of the repressor to $O_1$, $O_2$, and $O_3$ obtained from Equation (1) with the data of reference 14 for the repression levels ($R_{O_m}$) of a *lac* promoter with a single binding site $O_m$ with the sequence of $O_1$, $O_2$, or $O_3$. The number of repressors $N$ was increased over the wild type level ($N \simeq 10$) to averages of about 50 and 900 per cell. $\langle \Delta G_{O_m} \rangle$ is the average free energy for the two different numbers of repressors per cell. Explicitly, $\Delta G_{O_m} = -\ln\left(\frac{R_{O_m} - 1}{N}\right)$ and $\langle \Delta G_{O_m} \rangle = \frac{1}{2}\left(\Delta G_{O_m}(N=50) + \Delta G_{O_m}(N=900)\right)$. The units of energy are $k_B T$, with $k_B$ the Boltzmann constant and $T$ the absolute temperature. The different strains of reference 14 were constructed as follows. Plasmids with the *lacZ* gene under the control of the natural lac promoter and the three lac operators ($O_1$, $O_2$, and $O_3$) were integrated into the chromosome of a strain lacking the *lacZ* and *lacI* genes (BMH 8117 F'). The operators were either unchanged or altered by site-directed mutagenesis. The Lac repressor was expressed from a plasmid. (For more details see reference 14 and references therein.)

**TABLE 2:** Looping contribution to the free energy, $\Delta G_l$, for a *lac* promoter with a main operator (with the sequence of $O_1$, $O_2$, or $O_3$) and an auxiliary operator (with the sequence of $O_2$) obtained from Equation (3) with the average values of Table 1

($\Delta G_{O_1} = \langle \Delta G_{O_1} \rangle$, $\Delta G_{O_2} = \langle \Delta G_{O_2} \rangle$, and $\Delta G_{O_3} = \langle \Delta G_{O_3} \rangle$) and the data of reference 14.

Explicitly, $\Delta G_l = -\ln\left(\dfrac{(R_{O_m}-1)\left(1+Ne^{-\langle \Delta G_{O_a} \rangle}\right) - Ne^{-\langle \Delta G_{O_m} \rangle} - N(N-1)e^{-(\langle \Delta G_{O_m} \rangle + \langle \Delta G_{O_a} \rangle)}}{Ne^{-(\langle \Delta G_{O_m} \rangle + \langle \Delta G_{O_a} \rangle)}}\right)$.

To check the consistency of the results, $R_{O_m}$ was computed [Equation (3)] for all the cases considering $\Delta G_l = \langle \Delta G_l \rangle = -6.39$; i.e.,

$R_{O_m} = 1 + \dfrac{Ne^{-\langle \Delta G_{O_m} \rangle} + Ne^{-(\langle \Delta G_{O_m} \rangle + \langle \Delta G_{O_a} \rangle + \langle \Delta G_l \rangle)} + N(N-1)e^{-(\langle \Delta G_{O_m} \rangle + \langle \Delta G_{O_a} \rangle)}}{1 + Ne^{-\langle \Delta G_{O_a} \rangle}}$. The units of

energy and the experimental details of reference 14 are as in the caption of Table 1.

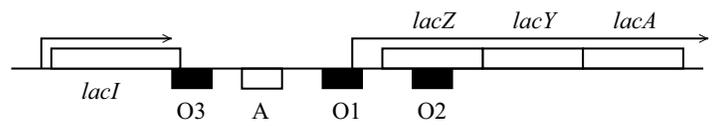

FIGURE 1

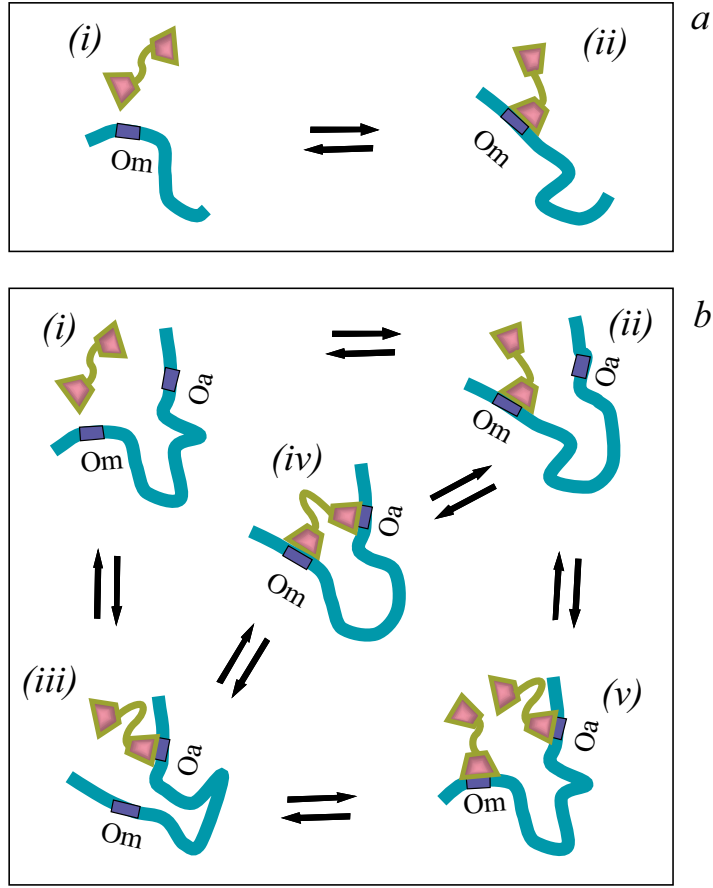

FIGURE 2

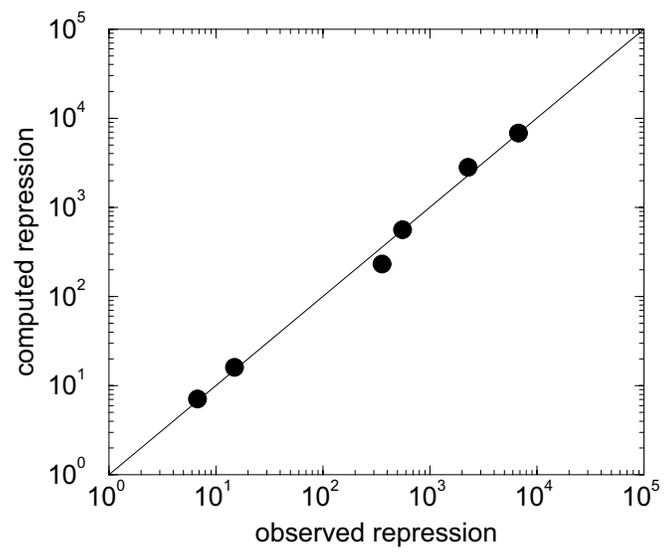

FIGURE 3

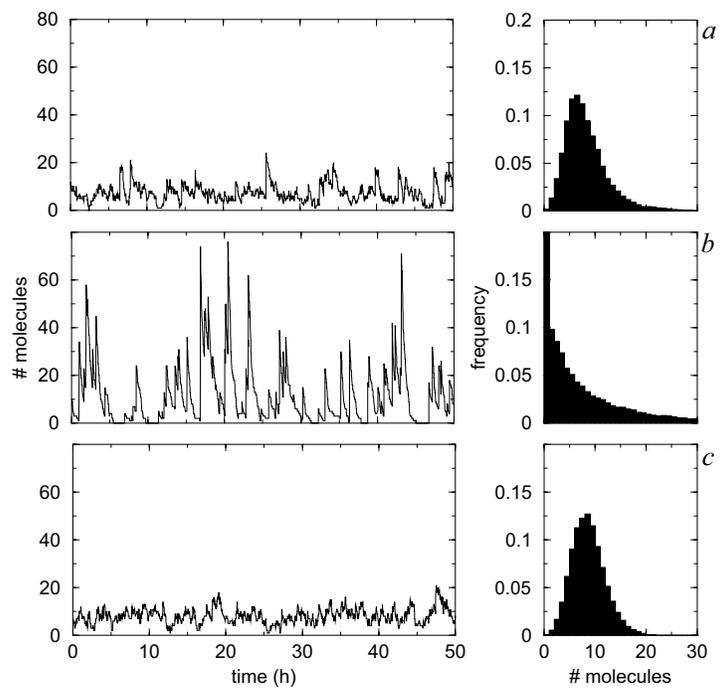

FIGURE 4

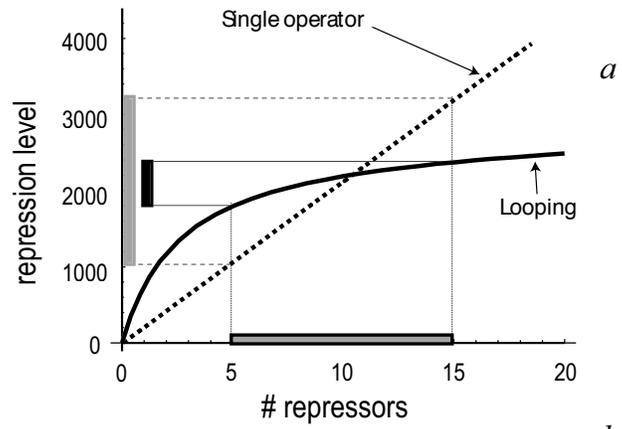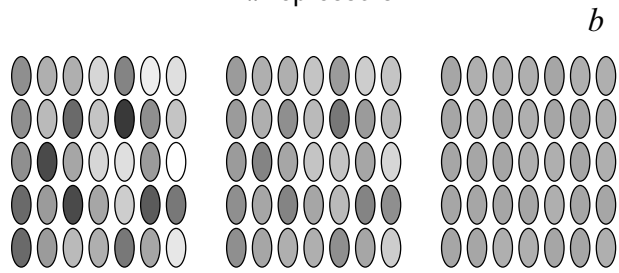

FIGURE 5

| $O_m$ | $N$ | Repression level[14] | $\Delta G_{O_m}$ | $\langle \Delta G_{O_m} \rangle$ |
|---|---|---|---|---|
| $O_1$ | 50 | 200 | $-1.38$ | $-1.52$ |
|  | 900 | $4,700$ | $-1.65$ |  |
| $O_2$ | 50 | 21 | 0.92 | 0.98 |
|  | 900 | 320 | 1.04 |  |
| $O_3$ | 50 | 1.3 | 5.12 | 4.61 |
|  | 900 | 16 | 4.09 |  |

TABLE 1

| $O_m - O_a$ | $N$ | Repression level[14] | $\Delta G_l$ | $R_{O_m}$ |
|---|---|---|---|---|
| $O_1 - O_2$ | 50 | 2,300 | −6.17 | 2,804 |
| $O_1 - O_2$ | 900 | 6,800 | −6.39 | 6,807 |
| $O_2 - O_2$ | 50 | 360 | −6.86 | 232 |
| $O_2 - O_2$ | 900 | 560 | −6.38 | 563 |
| $O_3 - O_2$ | 50 | 6.8 | −6.33 | 7.1 |
| $O_3 - O_2$ | 900 | 15 | −6.22 | 16 |

**TABLE 2**